\documentclass[twocolumn,showpacs,showkeys,prc,aps]{revtex4}
\usepackage{epsfig}
\newcommand{ \be }{\begin{equation}}      
\newcommand{ \ee }{\end{equation}}
\newcommand{ \bea }{\begin{eqnarray}}
\newcommand{ \eea }{\end{eqnarray}}

\newcommand{ \vs } {{\vec{s} }}
\newcommand{ \vb } {{\vec{b} }}

\newcommand{ \npart } {{N_{N-part}}}

\newcommand{ \ncoll } {{N_{N-coll}}}

\begin{document}

\title {
Nucleon participants or quark participants?
}


\author{S. Eremin$^{1,2}$, and S. Voloshin$^1$}
\affiliation{$^1$Wayne State University, 666 W. Hancock, Detroit, MI
48201, USA
\\
$^2$Moscow Engineering Physics Institute, 31 Kashirskoe shosse, 
Moscow, 115409, Russia}

\begin{abstract}
We show that centrality dependence of charged particle pseudorapidity
density at midrapidity in Au+Au collisions at RHIC 
is well described as proportional to the
number of participating constituent quarks. 
In this approach there is no need for 
an additional contribution from hard processes usually considered
in the models based on the number of the nucleon participants.
\end{abstract}

%\date{\today, version 10}
\date{March 18, 2003}
\maketitle

\section{Introduction}

Charged particle multiplicity densities near mid rapidity in high energy
nuclear collisions depend strongly on collision centrality. 
In order to better understand this dependence and to
disentangle pure nuclear effects, this density is often plotted as per
nucleon participant pair~\cite{phobos}.  
Participants are the nucleons that have encountered at least 
one inelastic collision. 
The number of participants at a given centrality is usually calculated 
in the Glauber model either
analytically or with the help of a Monte-Carlo algorithm. 
The charged particle density per participant 
increases with centrality.
In the 5\% most central Au+Au collisions it is about 20\% larger
than that in semi-peripheral collisions (50--70\% centrality region),
and it is about 50\% larger compared to $pp$ collisions at the same 
energy.

The reason for this  increase in number of produced particles
 per participant at midrapidity is still not well understood. 
The most common explanation of the phenomena involves  
particle production in hard processes. 
Hard processes have much smaller cross-sections than soft collisions 
and depend differently on collision centrality.
They scale with the number of binary 
collisions (the number of collisions the incident nucleon 
would experience if it were not 
altered at all while passing through the nucleus).  
The number of binary collisions increases with
centrality faster than the number of participants; 
this results in an increase
of particle production per participant nucleon 
as centrality increases.  
In such approaches the particle density is often presented 
simply as~\cite{kharzeev}:
\be
\frac{dN_{ch}}{dy} \; \propto \; \alpha \npart + (1-\alpha)\ncoll \,,
\ee
where the parameter $\alpha$ is the relative fraction 
of particles produced in the soft collisions, 
and $(1-\alpha)$ is the relative fraction 
produced in hard collisions.
With proper parameters, this fits the data fairly well;
see Ref.~\cite{phobos}.
Note, however, that in such models,
the relative contribution of hard processes is expected to increase with
the collision energy. 
The data seem inconsistent with such an energy dependence. 
	
In the approach proposed in this paper, 
both nuclei and single nucleons are considered as 
a superposition of constituent quarks (also often called as 
 ``dressed'' quarks or valons); 
there are three such dressed quarks per nucleon.
The concept of the constituent quarks has been known for many 
years (see Refs.~\cite{constq-first,hwa1982} and references therein).
The constituent quark approach is able to explain many features of
hadron-hadron and hadron-nucleus collisions~\cite{constq-kniga}. 
QCD calculations support the statement that inside a nucleon
there are three objects of the size of 0.1 -- 0.3~fm (see
Ref.~\cite{constq-shuryak} and references therein);
 for some recent works using or discussing 
constituent quarks, see also Refs.~\cite{constq-recent,hwa2002}.

In the constituent quark picture, a  $NN$ collision looks like a
collision of two light nuclei. Most often only 
one $qq$ pair interacts, with other quarks being {\it spectators}.
Only part of the entire nucleon energy is spent for particle
production at midrapidity (as $\sqrt{s_{qq}} \sim \sqrt{s_{NN}}/3$).
The quark spectators form hadrons in the nucleon fragmentation region.
In the case of $AA$ collisions, more than one quark per nucleon 
interacts due to the large nucleus size and the possibility for quarks from 
the same projectile nucleon to interact with different target nucleons. 
The goal of this study is to find the number of produced particles per
participant quark (pair) and to check for its centrality dependence.  

\section{Calculations of the number of participants}   %%%%%%%%%%%%%%%%%%%%%%%

We calculate mean number of nucleon/quark participants 
using a Monte-Carlo based implementation~\cite{misko}
of the nuclear overlap model \cite{nom}. 
We use Woods-Saxon nuclear density profile
\be
n_A(r)=\frac{n_0}{1+\exp [(r-R)/d]},
\ee
with parameters: $n_0=0.17\;{\rm fm}^{-3},
R= (1.12 A^{1/3} - 0.86^{1/3})~{\rm fm}, 
d=0.54~{\rm fm}.$

In the nuclear overlap model, 
the mean number of participants in $A+B$ collision at impact 
parameter $b$ is given by
\bea
&&N_{part,AB}=
\int d^2s \; T_A(\vs)\{1- [1-\frac{\sigma_{NN} T_B(\vs -\vb)}{B}]^B \}
\nonumber \\
&&+
\int d^2s \; T_B(\vs)\{1- [1-\frac{\sigma_{NN} T_A(\vs -\vb)}{A}]^A \},
\eea
where 
$ T(b) =\int_{-\infty}^{\infty} dz\; n_A(\sqrt{b^2+z^2})$ is the
thickness function; then  $(1-\sigma_{NN} T_A(b)/A)^A$ is the
probability for a nucleon to pass through the nucleus without any collision.
We use the inelastic 
$NN$ cross section $\sigma_{NN}=41$~mb  at $\sqrt{s_{NN}}=130$~GeV. 

The number of participating nucleons for a given centrality
can be determined directly using the web interface~\cite{misko}.
In order to calculate the number of participating quarks we
downloaded { FORTRAN} code and modified it by
increasing the density three times
($n_0^q = 3 n_0 = 51~{\rm fm}^{-3}$) 
and changing  $\sigma_{NN}$ to  $\sigma_{qq}$.
For the quark-quark cross-section in our calculation, we use two values
$\sigma_{qq}=\sigma_{NN}/9=4.56$~mb and a somewhat arbitrary value of
$\sigma_{qq}=6$~mb; the latter was  used mostly to illustrate 
the sensitivity of the
results to the value of $\sigma_{qq}$. 
The choice of  $\sigma_{qq}=6$~mb 
is not unreasonable since at RHIC energies,
approximately 1.2 -- 1.3 quarks per nucleon can participate in a single
$NN$-collision~\cite{constq-kniga}. 
In principle, $\sigma_{qq}$ could be probably as high as 8~mb based on 
the early estimates of $(r_q/R_N)^2 \sim 1/5$~\cite{hwa1982}.

\begin{figure}[htbl]
\hspace*{-5mm}
\includegraphics[height=8.5cm]{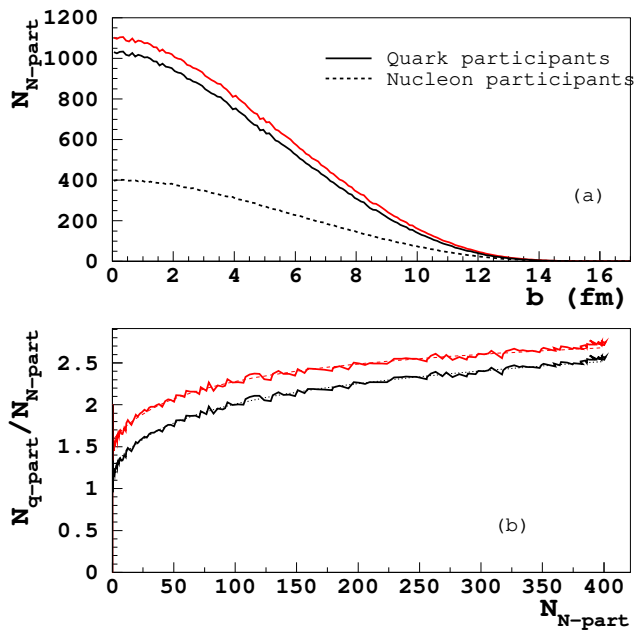}
\caption{Impact parameter dependence of  (a) the number of the nucleon and
the quark participants, and (b) the ratio of $N_{q-part}/N_{N-part}$.
The quark participant curves are shown for $\sigma_{qq}=4.56$~mb
(lower) and 6~mb (upper curve).  
}
\label{fig:npartvsb}
\end{figure}

Fig.~\ref{fig:npartvsb}a shows the number of
the nucleon and quark participants vs. impact parameter. 
Fig.~\ref{fig:npartvsb}b presents the centrality dependence 
of the ratio of $N_{q-part}/N_{N-part}$. 
Smooth curves are the polynomial fits to the Monte-Carlo results 
to smooth out the statistical fluctuations. 
The ratio  $N_{q-part}/N_{N-part}$ is used later for 
the renormalization of the particle pseudorapidity 
distributions from per nucleon participant to per quark participant. 

\begin{figure}[htbl]
\hspace*{-5mm}
\includegraphics[height=6.cm]{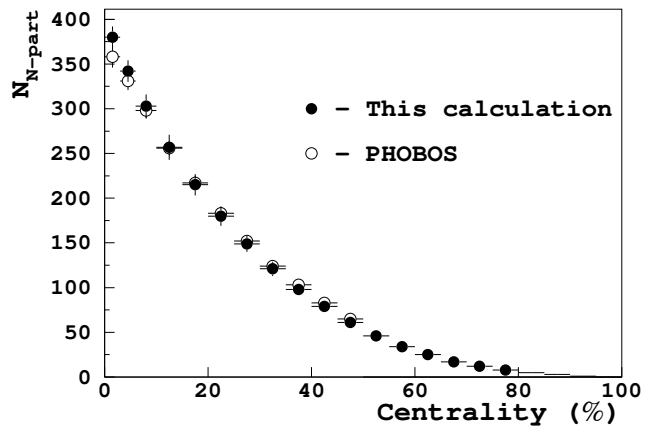}
\caption{Mean number of nucleon
participants {\it vs} centrality in the nuclear overlap model (solid symbols)
and from PHOBOS calculations (open circles).}
\label{fig:npartvscent}
\end{figure}

Fig.~\ref{fig:npartvscent} presents the comparison of our calculation 
of $N_{N-part}$ in the nuclear overlap model 
with PHOBOS calculations~\cite{phobos} based on the HIJING 
model.
The number of participants is plotted as a function of a given fraction
of the total cross section.
Open circles represent PHOBOS calculations.  
The nuclear overlap model results (using the Woods-Saxon density profile, 
the same as used in the HIJING model) are shown by solid symbols. 
Note a small deviation of our calculations from that of PHOBOS 
in very central region; this is not important for the conclusion 
of the current study. 

The PHOBOS Collaboration presents their results on centrality dependence of
the charged particle pseudorapidity density by plotting it vs. the number 
of the nucleon participant pairs. 
In this paper, we continue to use the same quantity 
for the centrality characterization,
but note that in the constituent quark picture, 
the number of the nucleon participants no longer 
has the meaning of the number of
the particle production sources.

\section{Results}  %%%%%%%%%%%%%%%%%%%%%%%%%%%%%%%%%%%%%%%%%%%%%%%%

\begin{figure}
\hspace*{-5mm}
\includegraphics[height=7.5cm]{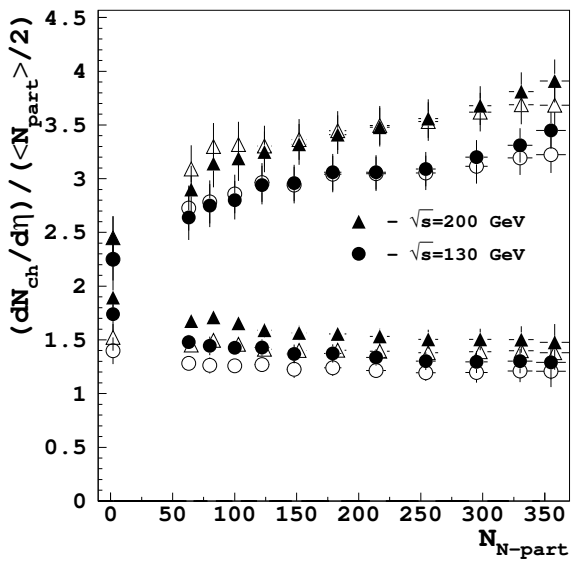}
\caption{$N_{ch}$ per nucleon and quark participant pair {\it vs} 
centrality. The results for quark participant pair are shown 
for $\sigma_{qq}=4.56$~mb (solid symbols) and $\sigma_{qq}=6$~mb (open
symbols). 
}
\label{fig:multpernpart}
\end{figure}

The centrality dependence of the charged particle 
multiplicity per participant pair 
is shown in Fig.~\ref{fig:multpernpart}. 
The results per nucleon participant pair are in the upper part of the
figure and the results per quark participant pair are shown in the
lower part. 
The original PHOBOS results~\cite{phobos} on $dN_{ch}/d\eta$ 
per nucleon participant pair (calculated using HIJING) 
for $\sqrt{s_{NN}}=130$~GeV and 200~GeV are shown in solid symbols. 
In open symbols we also show 
the same results renormalized for the number of the nucleon participant 
pairs from our calculations using the nuclear overlap model. 
Our main result -- the centrality dependence of the charged particle 
pseudorapidity density per quark participant pair -- is presented in 
the same plot.

We observe no, or even slightly decreasing, dependence of 
$(dN_{ch}/dy) / N_{q-part}$ on centrality,
with the ratio being dependent only on the
energy of the collision.
The slight decrease in particle production at midrapidity with 
increasing centrality could be due either to low values of 
the constituent quark inelastic cross section used in our calculations
or to parton saturation effects.    
Note that in the constituent quark picture,
$(dN_{ch}/dy) / N_{q-part}$ as a function of centrality depends very
weakly on the collision energy, as the change in the inelastic cross
section is probably less than 5\% between $\sqrt{s_{NN}}=130$~GeV and
200~GeV.    

\begin{figure}
\hspace*{-5mm}
\includegraphics[height=7.0cm]{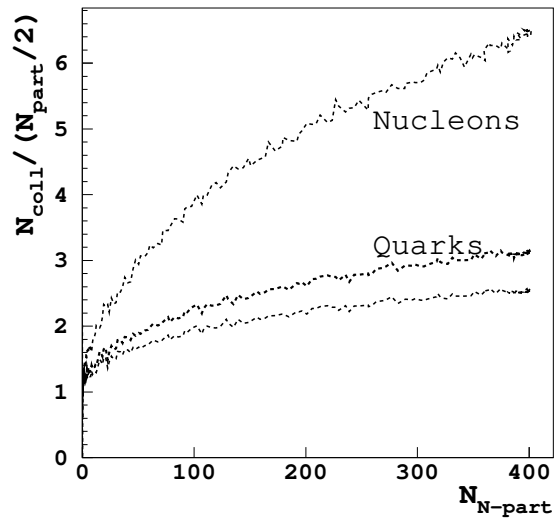}
\caption{
Centrality dependence of the ratio of binary collisions 
to the number of nucleon and quark (upper curve corresponds to
$\sigma_{qq}=6$~mb and lower to $\sigma_{qq}=4.56$~mb) participants.
}
\label{fig:nbin}
\end{figure}
Hard processes scale with the number of binary collisions.
Although it was not necessary to include the contribution of hard processes
into our calculation in order to describe the centrality dependence of 
the charged particle density at midrapidity, we have calculated 
the number of binary collisions as well: see Fig.~\ref{fig:nbin}.  
Note that the number of binary collisions per participant has a 
much weaker centrality dependence in the constituent quark approach 
than it has in the the nucleon participant model.

\section{Summary}

We have shown that the particle multiplicity density near mid rapidity 
scales linearly with the number of constituent quark participant pairs. 
The experimentally observed increase of $dN_{ch}/d\eta$ 
per nucleon participant pair with centrality
in this picture is explained by the relative increase 
in the number of interacting constituent quarks in more central collisions.
%As constituent quark cross section changes weakly with collision
%energy, $dN_{ch}/d\eta$ centrality dependence appears to be very similar for
%different collision energies. 

We thank D.~Mi\`{s}kowiec for making his code available to us.
This work was supported in part by 
U.S. DOE  Grant No. DE-FG02-92ER40713 and Wayne State University
Office of the Provost, Global Education Grant.

\end{document}